\documentclass[9pt,onecolumn,twoside]{revtex4}
\usepackage{graphicx}

\begin{document}
	
	\preprint{APS/123-QED}
	
	\title{Analysis of the phase-locking dynamics of a III-V-on-silicon frequency comb laser}
	
	\author{A. Verschelde}
	\author{M. Giudici}%
	\author{G. Huyet}
	\author{M. Marconi}
	\affiliation{Université Côte d\textquoteright Azur, Centre National de La Recherche Scientifique, Institut de Physique de Nice, F-06560 Valbonne, France
		}%
	
	
	\author{K. Van Gasse}
	\author{B. Kuyken}
	\affiliation{
		Photonics Research Group, Ghent University, Gent, Oost-Vlaanderen Belgium 9052
	}%
	
	\collaboration{CLEO Collaboration}
	
	\date{\today}
	
	\begin{abstract}
		We present the detailed characterization of the phase dynamics of a telecom hybrid III-V-on-silicon passively mode-locked laser with a ring cavity. We explore the various regimes of operation as a function of gain current and saturable absorber bias voltage. We use a stepped-heterodyne measurement to quantify the spectral chirp and reconstruct the pulse envelop. With this technique we are able to identify regimes of near-transform-limited operation and we assess the degradation of mode-locking in the comb phase relationship when saturable absorber bias voltage is changed. Finally, we present a preliminary study of the phase-locking in hybrid mode-locked operation and demonstrate transform-limited pulses.
	\end{abstract}
	
	\maketitle

\section*{Introduction}
Passive Mode-Locking (PML) is a well-known technique that uses a saturable absorber (SA) inside a laser cavity to generate a pulsed laser output with a broad optical frequency comb. This technique has been employed in a variety of laser devices and for a large number of applications including high resolution imaging \cite{JBiomedOpt}, optical signal processing \cite{Vlachos} and time-resolved measurements \cite{Sansone}. Nowadays, extensive work is dedicated to the generation of Optical Frequency Combs (OFCs) with integrated semiconductor laser technologies to satisfy the requirements for low energy consumption, small footprint and reproductibility.
A challenge for integrated technologies is to offer the possibility to reach low Free Spectral Range (FSR) values for optical cavities. Such a property is for example required for spectroscopic applications \cite{dualcomb}, where a ultra-dense set of comb lines allows to obtain a high resolution in the spectral analysis of gas.
Former monolithic integrated technologies  showed large bandwidth OFC emission but the FSR could not reach values lower than 10 GHz due to the losses imposed by the III-V material \cite{moska}. This limitation has been recently bypassed by the use of hybrid laser technologies, where the optical cavity is made of a Si waveguide and light is evanescently coupled from the III-V active region to the underneath passive Si layer \cite{keyva,keyvaOL}. The low-loss Si waveguide enables the propagation of light in the cavity over several cms; OFC lasing in PML hybrid lasers has recently been demonstrated with 1 GHz FSR, providing a ultra-dense comb spectrum containing more than 1400 lines \cite{ultradense}. The hybrid III-V-on-Si laser technology has proved mature enough for spectroscpic applications \cite{vanGasse} and better understanding of the noise performance and pulse quality of these lasers is required. In PML semiconductor lasers, the overall performances are strongly affected by the SA properties, hence the ability to measure the influence of the SA during lasing is essential to control the laser operation.

\begin{figure}
	\centering
	\includegraphics[scale=0.42,angle=0,origin=c]{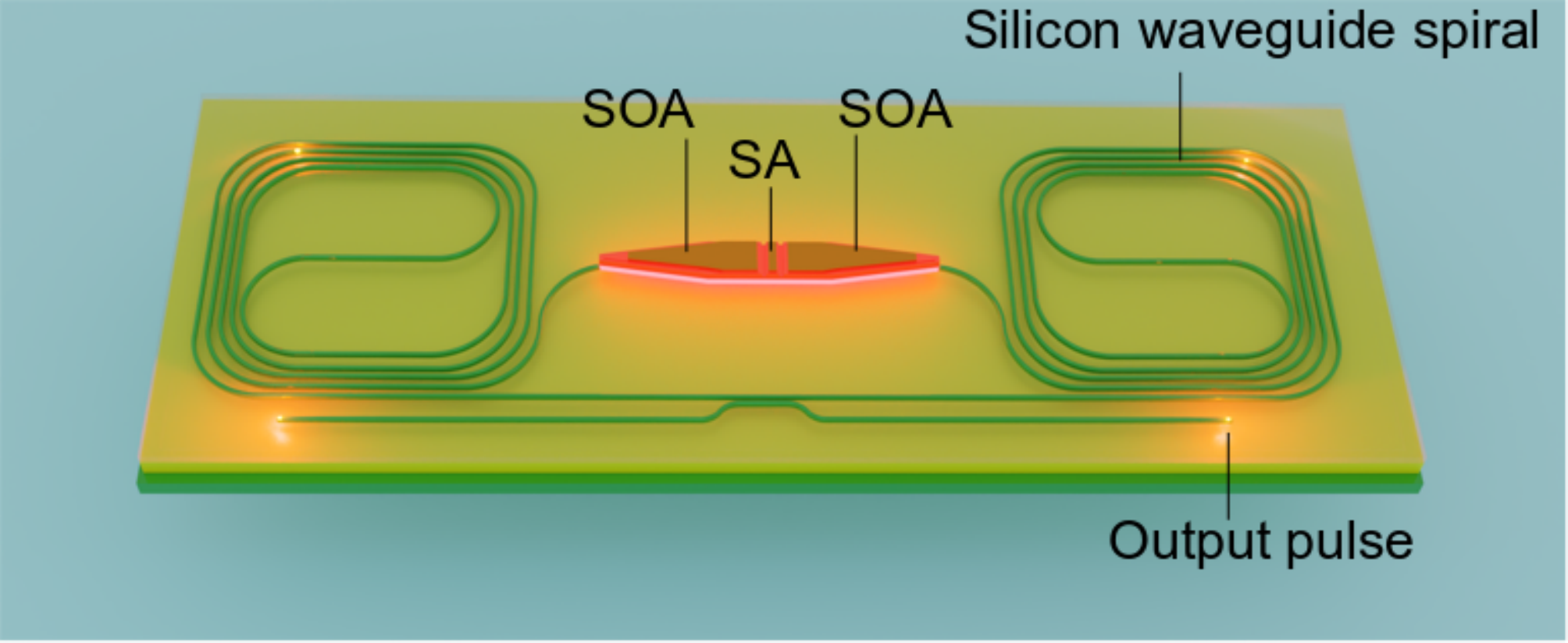}
	\caption{Layout of the hybrid 8 GHz ring frequency comb laser. Semiconductor Optical Amplifier (SOA) and Saturable Absorber (SA) are integrated within the active Indium Phosphide (InP) material. The long spiral waveguide is etched in the underlying Si membrane that is bonded to InP. Light is evanescently coupled between  active and passive layers.\label{fig:one}}
\end{figure}

In most of the relevant literature,  noise performances are typically well-assessed via the measurement of the fundamental RF tone and optical linewidths \cite{kef2008,habru}; on the other hand the pulse width is usually determined via a nonlinear autocorrelation setup \cite{Delfyett}, since typical timescales are too fast for current photodetectors technologies. However such a technique suffers severe limitations and drawbacks. First, the output of a nonlinear autocorrelation measurement is always symmetric even with non-symmetric pulses. Second, a nonlinear autocorrelation trace can correspond to several pulse envelops, an initial shape (gaussian, secant hyperbolic) must be inferred to determine the pulse width;  this technique is thus quite unreliable to precisely assess transform-limited (TL) operation from a mode-locked laser. Finally, this technique relies on a second or third order nonlinear susceptibility of a material and can thus be only applied to powerful enough laser outputs, which discards integrated low power pulsed laser sources. 
In this paper, we propose to assess the mode-locking performance and pulse formation in hybrid III-V-on-Si PML laser by monitoring the spectral phase of the frequency comb output. We therefore choose to use a Stepped-Heterodyne (SH) measurement to access phase and amplitude of each comb line which allows us to reconstruct the temporal envelop and phase chirp of the pulse with a temporal resolution smaller than the picosecond \cite{SH}. Such a technique was already used with monolithic PML lasers to show that no TL operation could be obtained with monolithic devices, and evidence the frequency chirp in the output comb \cite{moska}.
In this work we disclose and characterize regimes of near-TL pulsed operation with hybrid PML ring lasers and give for the first time a quantitative description of the degradation of the phase locking between the OFC lines when the SA bias voltage values depart from optimal ones.  Such a degradation manifests in the appearance of ripples in the reconstructed pulse envelop. We also perform an analysis of the phase dynamics in a regime of resonant hybrid (passive and active) mode-locking, where a RF modulation is applied on top of the SA bias voltage.

\begin{figure}
	\centering
	\includegraphics[scale=0.17,angle=0,origin=c]{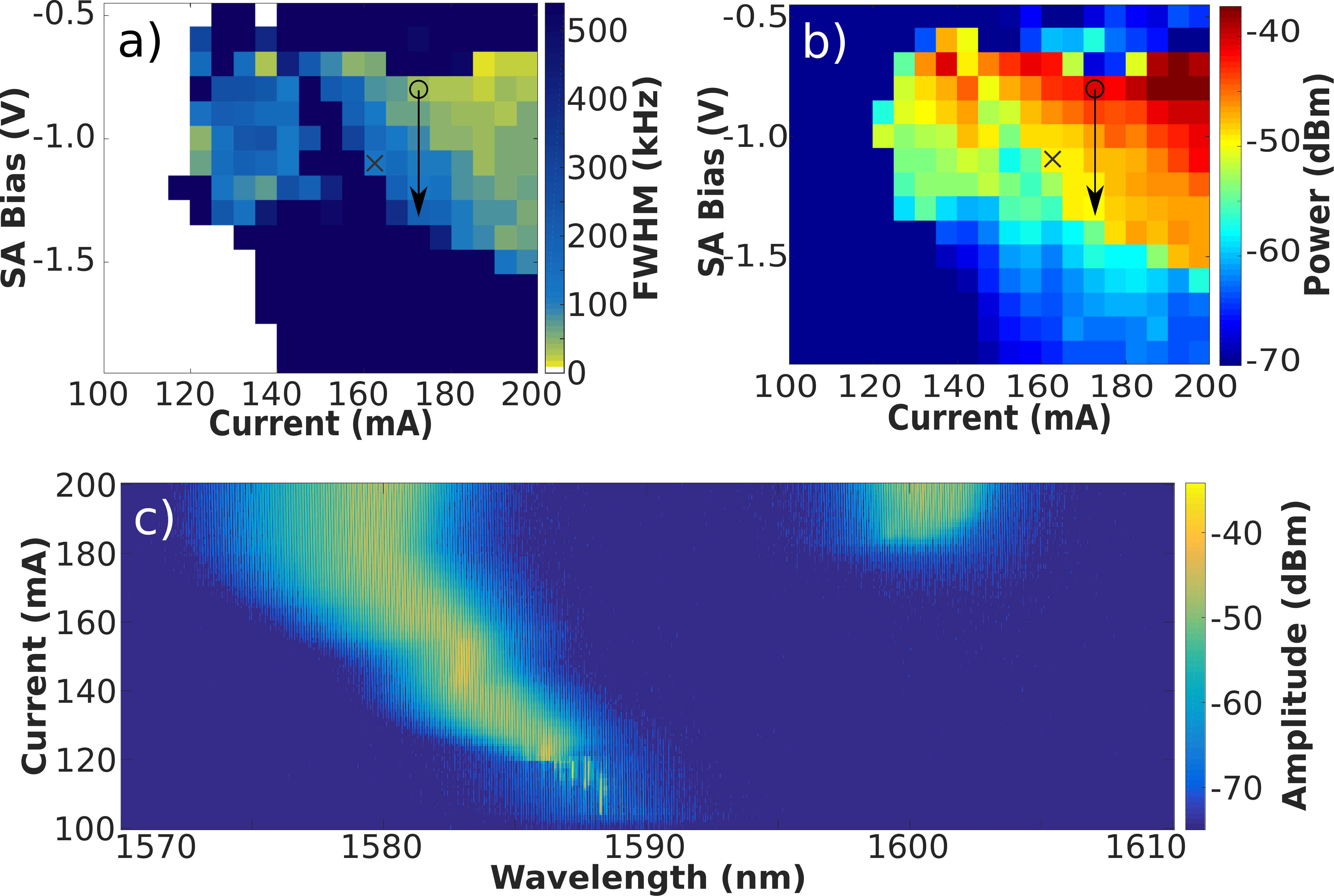}
	\caption{a,b) RF fundamental tone FWHM (a) and amplitude (b) as a function of $V_{SA}$ and pumping current I.  c) Optical spectra as a function of pumping current for $V_{SA}$ = -1 V. \label{fig:two}}
\end{figure}

\begin{table}
	\centering
	\caption{\bf RF fundamental tone FWHM and amplitude for different $V_{SA}$ and I = 170 mA in the PML regime.}
	\begin{tabular}{lccccr}
		\hline
		$V_{SA}$(V) & -0.8 & -0.9 & -1 & -1.1 & -1.2 \\
		\hline
		RF FWHM (kHz)  & 36 & 60 & 86 & 104 & 118   \\
		\hline
		RF peak (dB) & -40.8 & -43.9 & -46.9 & -47.6 & -49.1 \\
		\hline
	\end{tabular}
	\label{tab:1}
\end{table}	
	
The laser under study has a ring cavity configuration similar to the one presented in \cite{keyva} (Fig.\ref{fig:one}). The ring cavity allows to outcouple the two counterprapogating pulses to different waveguides in order to avoid the spurious tones identified with linear cavities close by DBRs \cite{keyva}. The FSR is 8 GHz, which corresponds to a cavity length of about 5 mm. The laser is temperature controlled and pumped with a low-noise current source. The typical average output power in the PML regime is 10 to 50 $\mu W$. This low value is due to the approximate -20 dB coupling loss from chip to free space, resulting in about 1 mW average power on the chip. In order to find the optimal PML operations for the laser, we first characterize the fundamental RF tone Full Width Half Maximum (FWHM) and amplitude. Fig. \ref{fig:two}a shows a map of the RF fundamental tone FWHM when the SA voltage and pumping current are varied. This map allows to identify a threshold current of 130 mA for the appearence of a peak in the RF spectrum, and a threshold for the SA reverse bias voltage of -0.6 V. The white region indicates that no RF tone is present, while the dark blue region corresponds to a RF FWHM larger than 500 kHz. Within these two limits we identify a coloured region where the RF FWHM varies between 20 and 150 kHz. The mode-locking quality is then further assessed by mapping the amplitude of the fundamental RF tone in the $I/V_{SA}$ plane. From these two maps we evidence a region of best PML operation comprised in the limit of 160 to 200 mA for the current and -0.8 to -1.3 V for $V_{SA}$.

\begin{figure}[t]
	\includegraphics[scale=0.135,angle=0,origin=c]{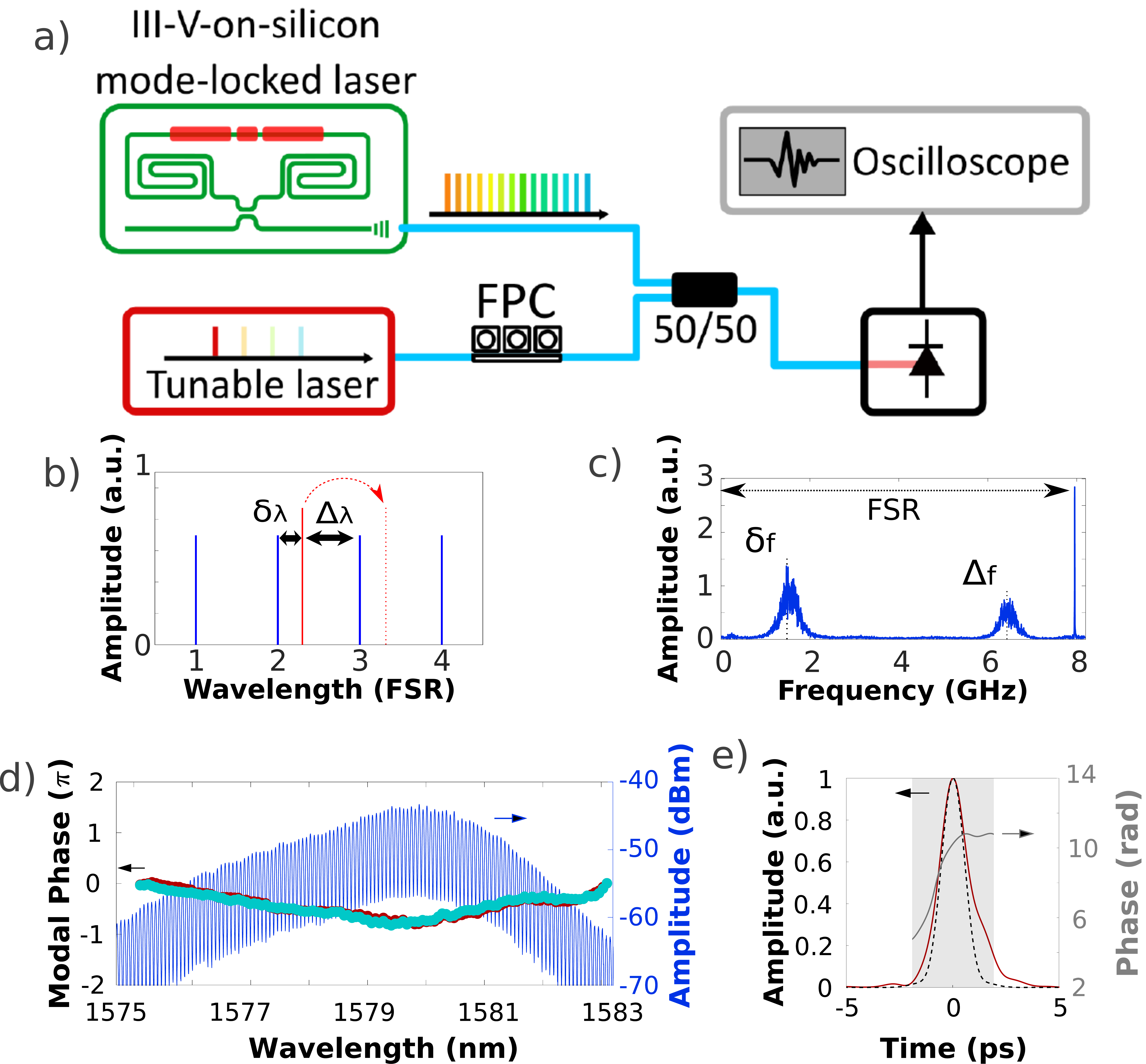}
	\caption{Near-transform-limited operation of the PML laser analyzed with the SH technique. a) Layout of SH measurement, FPC: fiber polarization controller. b) In SH, the tunable laser wavelength (red) is positioned between consecutive modes of the mode-locked laser. c) Fourier transform of a typical beating signal acquired  with a fast photodiode for one position of the tunable laser.  d) OSA trace (blue) and  modal phase (circles) for each comb line (red and blue circles stand for measurement 1 and 2 respectively). e) Reconstructed pulse envelop (red), temporal phase evolution (solid black line) for measurement 1. The black dotted line represents the simulated perfect TL pulse envelop. \label{fig:three}}
\end{figure}
Finally, Fig. \ref{fig:two}c shows the evolution of the optical spectrum obtained with a 10 pm resolution as a function of current for $V_{SA}$ = -1 V. One can evidence that a broad spectrum emerges from a multimode unstable operation at a current of 120 mA. As current is increased, the spectrum gets broader up to a value of 5 nm at -10 dB at 180 mA. We then notice the emergence of a second lobe in the spectrum that is red-detuned by 20 nm. Therefore, to avoid discontinuity in the analysed comb phase dynamics, we chose to avoid this region and remain in a current limit between 160 and 180 mA. From the presented data it is clear that the semiconductor PML laser shows non-trivial optical and RF features as a function of the parameters.  
	
We now intend to further analyse the characteristics of the phase-locking of spectral lines  during  PML operation. Using the SH technique, we are able to monitor the phase relationship between consecutive comb lines and further reconstruct the temporal envelop and phase evolution.
The SH technique is described in Fig. \ref{fig:three}a-c (see \cite{SH} for further details). This technique allows to obtain a complete phase spectrum by beating a CW laser with consecutive spectral lines of the hybrid laser comb. At each step the CW tunable laser wavelength is increased by one FSR of the comb laser, the beating signal with the nearest comb lines are acquired with a fast photodiode (Fig. \ref{fig:three}b,c). After filtering in the fourier space, we extract the three signals $S_\delta$, $S_{\Delta}$ and $S_\Omega$ respectively at the frequencies $\delta, \Delta = \Omega-\delta$ and $\Omega = FSR$. The following operation $S_\delta  S_{\Omega-\delta}  \bar{S_\Omega}$ is performed at each step in order to retrieve the phase difference between consecutive lines. The amplitude of each line is obtained from the integration of the RF power spectral density of the beating tones. 

\begin{figure}
	\includegraphics[scale=0.175,angle=0,origin=c]{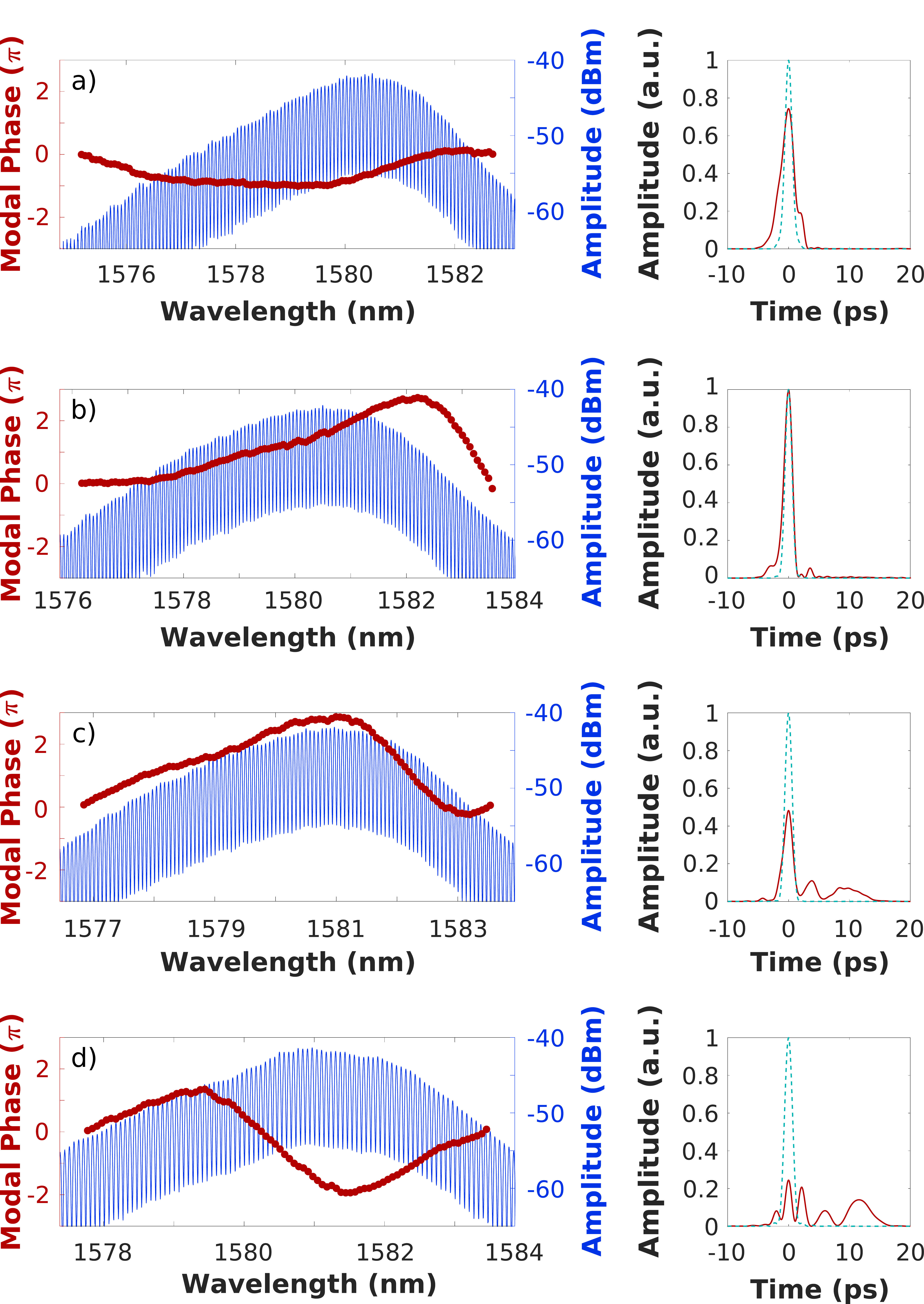}
	\caption{Phase-locking degradation (left) and pulse distortion (right) for increasing $V_{SA}$ values from the near-TL operation point, I = 170 mA. a) $V_{SA}$ = -0.9 V, b) $V_{SA}$ = -1 V, c) $V_{SA}$ = -1.1 V, d) $V_{SA}$ = -1.2 V. Blue dotted lines on the right panels stand for the simulated perfect TL pulses. \label{fig:four}}
\end{figure}
The phase and amplitude information of each comb line allows to reconstruct the temporal phase and pulse envelop. Fig. \ref{fig:three}d,e show the result of the SH measurement for the mode-locked operation referenced by the circle in the maps of Fig. \ref{fig:two}a,b (I = 170 mA, $V_{SA}$ = -0.8 V). This operation point corresponds to the best mode-locked situation in terms of RF FWHM (36 kHz) and amplitude (- 40.8 dB). The optical spectrum is shown in blue in Fig. \ref{fig:three}d. The corresponding phases of the comb lines are marked by circles. Red and light blue circles respectively stand for two consecutive phase measurements with the SH technique. As one measurement takes about one minute, this clearly shows that the phase dynamics is stable on a long timescale in our experiment while we perform the SH method.  The modal phase is almost linear with maximal deviation of only $0.5\pi$, hence group delay is almost constant and group delay dispersion is close to zero. This means that for this specific set of parameters,  the operation of the laser is near TL. Given that both the SA and amplifier will cause considerable chirp of the pulse, this means that for this point the chirp of the amplifier and SA almost exactly cancel. The reconstructed pulse envelop is shown in red in Fig. \ref{fig:three}e. To assess the quality of the TL operation, the pulse is compared with a perfect TL pulse that we have computed by setting numerically a constant value for the phase of each comb line. The TL pulse is symmetric and has a duration of 1.35 ps while the measured pulse is slightly asymmetric and its duration is 1.55 ps. Finally, Fig. \ref{fig:three}e shows the  phase evolution across the pulse. The slight departure from the TL operation can be witnessed from the curvature of the phase, which indicates that the instantaneous frequency (the derivative of the temporal phase) is not constant during the pulse.

The near-TL operation described above was obtained for a specific set of I and $V_{SA}$. We now use the SH technique to assess how phase-locking changes when increasing $V_{SA}$ for a fixed value of I = 170 mA as indicated by the arrow in Fig. \ref{fig:two}a,b. From Fig. \ref{fig:two}a,b we observe that this change in $V_{SA}$ causes an increase of the RF fundamental beat note linewidth and amplitude; the corresponding values are indicated in Tab. \ref{tab:1}. We now show the evolution of modal phase and pulse envelop  in Fig. \ref{fig:four}.   As we depart from the best TL operation, we see that small curvatures appear in the graphs of the modal phases, that induces a broadening and stronger ripples in the pulse envelop (Fig. \ref{fig:four}a).  The modal phase dispersion is even more pronounced in Fig. \ref{fig:four}b, however since the effect is localized on the side modes of the optical spectrum, the pulse envelop shows less evident rippling and broadening. In the cases c and d (resp. $V_{SA}$ = -1.1 and -1.2 V), the dispersion occurs in the central modes of the optical spectrum, therefore the pulse envelop becomes more affected. We would like to note that for each parameter, the measurement was performed five times in a row and each time the same evolution of the modal phase was measured. Therefore we can confidently state that the pulse envelop we reconstruct is not evolving on slow timescales as we perform the measurements. However the SH technique does not allow to state that fast changes of the modal phase do not occur in the course of the measurement.
\begin{figure}
	\includegraphics[scale=0.16,angle=0,origin=c]{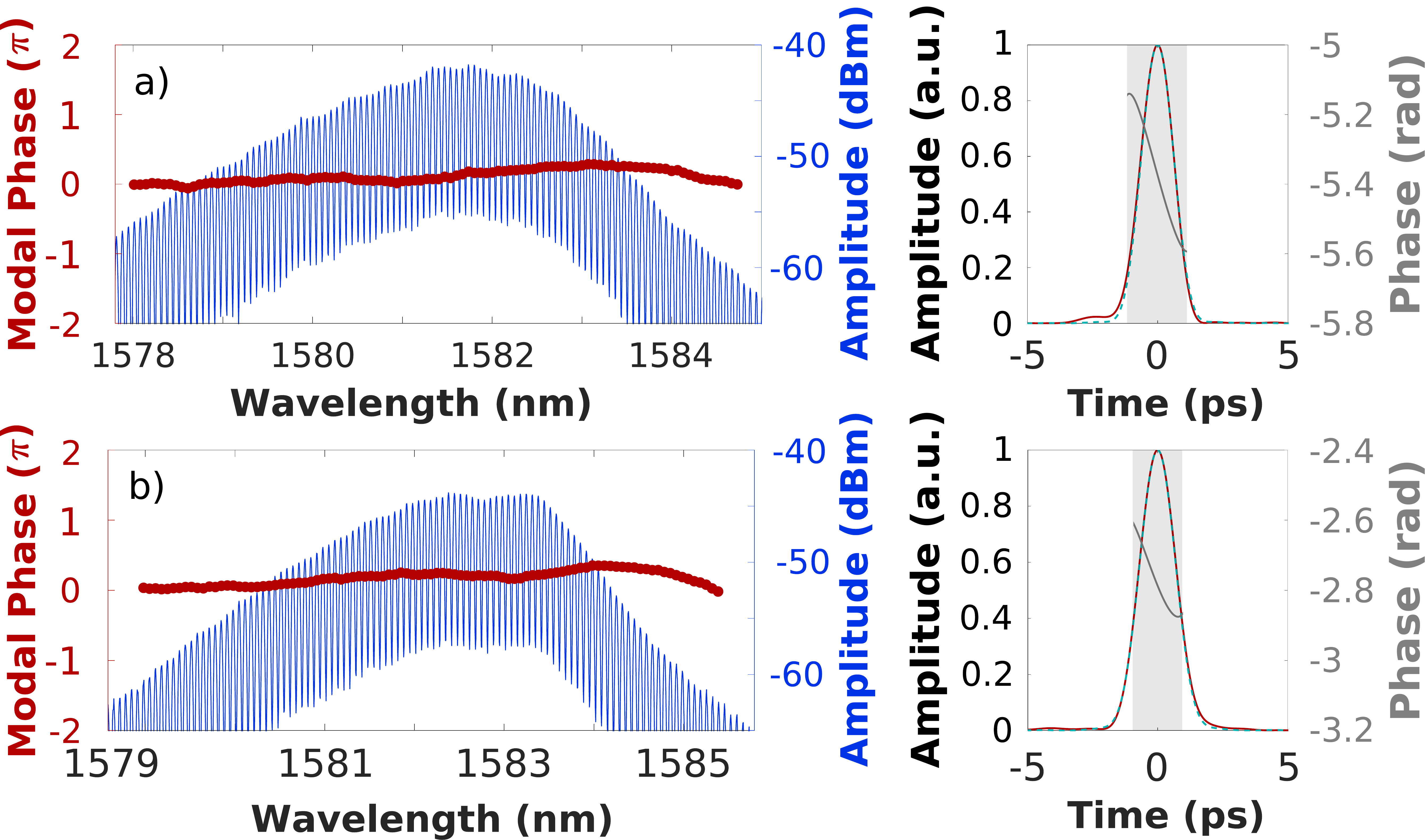}
	\caption{Modal phase, pulse envelop and temporal phase reconstruction in the hybrid mode-locking regime. a)  $V_{SA}$ = -1.2 V and I = 170 mA (same values as Fig. \ref{fig:four}d). b) $V_{SA}$ = -1.1 V and I = 160 mA.\label{fig:five}}
\end{figure}

So far, the pulse reconstruction has been performed while the laser was in a PML operation. However, the design of the ring lasers allows to apply a RF electrical modulation to the SA section and therefore operate the laser in the so-called hybrid mode-locking regime.  Such modulation has for example been used in \cite{vanGasse} in order to lock the repetition frequency of the laser to an external clock, and therefore avoid the effect of environnemental noise on pulse jittering.  We show in Fig. \ref{fig:five} how such modulation impacts the modal phase relationship and pulse envelop. In all cases, $V_{SA}$ consists in a 100 mV p-p sinusoidal signal added to a DC offset. The RF frequency is chosen such that it stands in the locking cone where the laser FSR will follow the electrical clock (see \cite{vanGasse} for details). In Fig. \ref{fig:five}a, we show the modal phase and pulse envelop when the DC offset and pump current are respectively -1.2 V and 170 mA (same parameters as Fig. \ref{fig:four}d). The measured pulse (red) is compared to a perfect TL pulse obtained by setting numerically all phases at the same value. We can observe that the modulation has allowed to flatten the modal phase with respect to the solitary PML operation shown in Fig. \ref{fig:four}d. As a consequence, the reconstructed pulse envelop matches now almost perfectly the optimal TL situation. This operation is also characterized by a linear phase evolution across the pulse, hence a constant instantaneous frequency. To demonstrate the robust effect of the RF electrical injection on the phase dynamics of the laser modes, we apply it for a different set of I/$V_{SA}$ where the operation of the PML laser is characterized by even larger values of RF FWHM and smaller amplitude when solitary. For this we only decrease the pump current value to 160 mA while keeping a fixed  $V_{SA}$ = -1.1 V (cross in Fig. \ref{fig:two}a,b). At this operation  point, RF FWHM and amplitude are respectively 122 kHz and -49.3 dB. We observe in Fig. \ref{fig:five}b that the RF modulation allows again to generate a perfect TL pulse with a linear phase evolution. Although hybrid mode-locking has been extensively demonstrated and studied, the details of operation are complex and not fully understood \cite{ultradense, vanGasse}. With the presented results, we show that the SH technique allows to investigate, in detail, the possibility of pulse shaping using the RF modulation of the SA.

In conclusion, we have presented the first stepped-heterodyne analysis of the mode-locking operation of a hybrid III-V-on-Si frequency comb laser. The SH technique provides information on the pulse envelop through the phase relationship between the comb lines, which is a clear advantage over nonlinear autocorrelation measurement. Moreover this technique does not require any pulse amplification and is therefore perfectly suited for the characterization of the frequency comb emitted by low power integrated laser sources. We have evidenced thanks to this technique a near-transform-limited operation characterised by a pulse duration of 1.55 ps and near-zero group delay dispersion. The SH technique has allowed to evidence how sensitive the phase-locking dynamics is with respect to the SA bias voltage. In fact, this work shows how deviations from best mode-locked operation in terms of RF linewidth enhancement manifests in the modal phase excursions, provocating rippling and broadening of the pulse envelop and complex temporal phase chirp. Finally, external-locking to a RF electrical clock being a simple and widely used method to stabilize a frequency comb, we have chosen to apply the SH technique in the hybrid mode-locking regime. We have measured perfect TL pulses with linear phase chirp when applying a sinusoidal modulation on top of the SA bias voltage that is resonant with the 8 GHz FSR of the laser cavity. Our analysis paves the way for further exploration, modeling and potential manipulation of the phase dynamics in semiconductor mode-locked lasers, with particular focus on the hybrid III-V-on-Si technologies where very long cavities can be fabricated \cite{ultradense, vanGasse}. These promising results can serve as a starting point for future investigations and would allow to identify new ways of controlling the pulse shape in semiconductor mode-locked lasers.


\section*{Acknowledgement}

European Research Council ELECTRIC. Flemish Research Council (FWO) Post-doctoral fellowship 12ZB520N. Conseil Régional Provence-Alpes-Côte d’Azur (Emplois Jeunes Doctorants 2018-2021, plateforme Optimal). 
	

\section*{References}	
\providecommand{\noopsort}[1]{}\providecommand{\singleletter}[1]{#1}%

\end{document}